\documentclass[showpacs,twocolumn]{revtex4}
\usepackage[dvips]{epsfig}
\newcommand{\LT}{\hat{\cal L}}
\newcommand{\clF}{{\cal F}}

\newcommand{\tM}{\tilde{M}}
\newcommand{\clW}{{\cal W}}
\newcommand{\clP}{{\cal P}}
\newcommand{\clC}{{\cal C}}

\newcommand{\rgl}{\rangle}
\newcommand{\lgl}{\langle}
\newcommand{\ep}{\epsilon}

\newcommand{\be}{\begin{equation}}
\newcommand{\ee}{\end{equation}}
\newcommand{\bea}{\begin{eqnarray}}
\newcommand{\eea}{\end{eqnarray}}
\begin{document}

\title{Lyapunov exponents in 1d disordered system with long-range
memory}

\author{Alexander Iomin}

\affiliation{Department of Physics, Technion, Haifa, 32000,
Israel}

\date{\today}
\begin{abstract}
The Lyapunov exponents for Anderson localization are studied in a one 
dimensional disordered system. A random Gaussian potential  with the 
power law decay $\sim 1/|x|^q$ of the correlation function is considered. 
The exponential growth of the moments of the eigenfunctions and their
derivative is obtained. Positive Lyapunov exponents, which
determine the asymptotic growth rate are found.
\end{abstract}


\maketitle

In this paper we consider Anderson localization
\cite{Anderson,Lee} in a one dimensional disordered system with a
long-range memory. The recent realization of disordered systems by
using ultra cold atoms \cite{G13,G14} in optical lattices and
microwave realization of the Hofstadter butterfly \cite{israilev}
show that the random potential in the experiments are highly
correlated. The increased interest in the problem of Anderson
localization in random potentials with long-range correlations is
also relevant to studies of the metal-insulator transition
\cite{garcia,moura}.

Anderson localization in a one dimensional disordered
system is described in the framework of the eigenvalue problem
\be\label{LR1} %
\ep\phi(x)=-\frac{d^2}{dx^2}\phi(x)-V(x)\phi(x)\, , %
\ee %
with a Gaussian random potential  $V(x)$. The long-range memory of
the disorder is modelled by the two point correlation function
$\clC(x)$ with the power law decay at the large scale
\be\label{LR2} %
\lgl V(x')V(x)\rgl=\clC_{q}(x-x')= \frac{C_{q}}{|x-x'|^{q}}\,
, %
\ee %
where $q>0$. It has been shown by various techniques that all
eigenfunctions are localized for correlated potentials with the
correlation decay rate  $0<q<1$ \cite{moura,LyraHavlin}. Spectral
properties of the random operator of Eq. (\ref{LR1}) (and its
discrete counterpart) were studied \cite{G7,G8,PF}. Due to the
physical interpretation, see discussion in Ref. \cite{garcia}, one
of the main results is the absence of the absolutely continuous
spectra for the random Schr\"odinger operator (\ref{LR1}) with the
correlation properties due to Eq. (\ref{LR2}). This means that the
eigenfunctions $\phi(x)$ are localized, and investigation of
Lyapunov exponents is a serious task related to localization of
the eigenfunctions.

The Lyapunov exponents are important in spectral theory, since
they govern the asymptotic behavior of the wave functions. They
are defined on the asymptotic behavior of the averaged envelope
$\gamma_s(\ep)=\lim_{x\to\infty}\frac{\lgl\ln\phi^2(x)\rgl}{2x}$.
It was shown by rigorous analysis that the positive Lyapunov
exponents are absent for the absolutely continuous spectrum, while
the positiveness of the Lyapunov exponents ensures that the
spectrum is pure point \cite{PF,LGP}.

In this paper, we calculate $\lgl\phi^2(x)\rgl $ of solutions of
Eq. (\ref{LR1}) for a certain energy $\ep$, with given boundary
conditions at some point, for example $\phi(x=0)$ and
$\phi^{\prime}(x=0)$, where prime means the derivative with
respect to $x$. Since the distribution of random potentials is
translationally invariant,  it is independent of the choice of the
initial point as $x=0$. It will be shown that this quantity grows
exponentially with the rate $\gamma(\ep)=\lim_{x\to\infty}\frac{
\ln\lgl\phi^2(x)\rgl}{x}>0$. Note that it is different from
$\gamma_s$, which supposes a knowledge of all the even moments
\cite{kirone,schomerus,pikovsky,fik}.

We develop a general procedure which is suitable  for calculation
of all moments of the wave function and its first derivative. To
this end the Schr\"odinger equation (\ref{LR1}) is considered as
the Langevin equation and the $x$ coordinate  as a formal time.
For the $\delta$ correlated process it can be easily mapped on the
Fokker-Planck (diffusion) equation for the probability
distribution function $\clP(\phi,\phi')$ \cite{LGP,halperin}.
Unlike this, the two point correlation function (\ref{LR2}), which
corresponds to the stationary process, leads to additional
integration over the formal ''time'' with a memory kernel. The
method of consideration enables one to observe the exponential
growth of $\lgl\phi^2(x)\rgl $ with the Lyapunov exponent
$\gamma(\ep)>0$.

Since the Schr\"odinger equation (\ref{LR1}) is a linear
stochastic equation, equations for the $2n$ moments of the type
\be\label{LR3} %
M_{k,l}(x)=\lgl [\phi(x)]^k[\phi '(x)]^l\rgl\, ,
~~k+l=2n, ~~k,l=0,1,2,\dots\, ,  %
\ee %
can be obtained in the closed form. To this end we rewrite Eq.
(\ref{LR1}) in the form of the Langevin equation. The $x$
coordinate is considered as a formal time on the half axis
$x\equiv\tau,~~\tau\in[0,\infty)$ and the new dynamical variables
$u(\tau)=\phi(x),~v(\tau)=\dot{u}=\phi'(x)$ are defined. In the
new variables the Langevin equation reads
\be\label{LR4} %
\dot{u}=v\, ,~~\dot{v}=-[\ep+V(\tau)]u\, , %
\ee %
where $V(\tau)$ is now the long-range correlated noise
\be\label{LR5} %
\clC_{\alpha}(\tau)=\frac{C_{\alpha}}{\tau^{1+\alpha}}\, . %
\ee %
It is convenient to set $q=1+\alpha$ and $C_q\equiv
C_{\alpha}$. In the new variables the expectation values of Eq.
(\ref{LR3}) are now $M_{k,l}(\tau)=\lgl u^kv^l\rgl$. Solutions of
Eq. (\ref{LR4}) are obtained as functionals
\be\label{LR6} %
v(t)=-\int_0^t[\ep+V(\tau)]u(\tau)d\tau\, ,~~~u(t)=\int_0^tv(\tau)d\tau\,
.
\ee %
Following \cite{pikovsky} we obtain a temporal equation for the
moments from the Langevin equation (\ref{LR4}) and its solutions
(\ref{LR6}). Differentiating $M_{k.l}(\tau)$ with respect to
$\tau$, we obtain
\be\label{LR7} %
\dot{M}_{k,l}=kM_{k-1,l+1}-l\ep M_{k+1,l-1}-l\lgl
V(t)u^{k+1}v^{l-1}\rgl\, .  %
\ee  %
The application of the Furutsu-Novikov formula \cite{klyatskin} to
the last term in Eq. (\ref{LR7}) yields
\bea\label{LR8} %
 &\lgl V(t)\clF[V(t)]\rgl=\frac{1}{2}\int_0^td\tau'\lgl V(t)V(\tau')\rgl
\Big\lgl\frac{\delta\clF[V(\tau)]}{\delta V(\tau')}\Big\rgl
\nonumber \\
&= -\frac{1}{2}(l-1)\int_0^t\clC_{\alpha}(t-\tau)M_{k+2,l-2}(\tau)d\tau \, .
\eea %
Here the solution of Eq. (\ref{LR6}) is used to obtain the
functional derivative of the functional
$\clF[V(\tau)]=u^{k+1}v^{l-1}$. Substituting the solution of Eq.
(\ref{LR8}) in Eq. (\ref{LR7}), we obtain that the temporal
behavior of the moments is described by the
fractional--differential equation
\be\label{LR9} %
\dot{M}_{k,l}=kM_{k-1,l+1}-l\ep
M_{k+1,l-1}+\frac{1}{2}l(l-1)D_t^{\alpha}M_{k+2,l-2}\, , %
\ee %
where the convolution integral in Eq. (\ref{LR8}) is the
fractional derivative $D_t^{\alpha}f(t)$
\be\label{LR10}%
D_t^{\alpha}f(t)=C_{\alpha}\int_0^t
\frac{f(\tau)d\tau}{(t-\tau)^{1+\alpha}}\,
.\ee %
Here the correlation function $\clC_{\alpha}(t)$ defines the
memory kernel, or  the causal function. Eqs. (\ref{LR9}) and
(\ref{LR10}) are relevant to the fractional Fokker-Planck
equations which  describe a variety of physical processes related
to fractional diffusion \cite{bouchaud,klafter,zaslavsky}. An
important technique for the treatment of the fractional equation
is the Laplace transform. It is worth stressing that both
analytical properties of this fractional integration and the
Laplace transform depend on $\alpha$.

For $-1<\alpha<0$ Eq. (\ref{LR9}) is readily solved by means of
the Laplace transform. Defining $\LT [M_{k,l}(t)]=\tM_{k,l}(s)$,
one obtains from Eq. (\ref{LR10}) $\LT[D_t^{\alpha}M_{k,l}(t)]
=C_{\alpha}\Gamma(-\alpha)s^{\alpha}\tM_{k,l}(s)$, where
$\Gamma(\alpha)$ is the gamma function. For simplicity,
disregarding the sign of the correlation function (\ref{LR5}), we
set $C_{\alpha}=2/\Gamma(-\alpha)$. Then, we introduce
$2n+1$-dimensional vectors ${\bf
M}_n(t)=\Big(M_{2n,0},M_{2n-1,1},\dots\, ,
M_{1,2n-1},M_{0,2n}\Big)$ in the ``time'' space and ${\bf
\tM}_n(s)=\LT[{\bf M}_n(t)]$ in the Laplace space,
correspondingly. Then the solution of Eq. (\ref{LR9}) is the
Laplace inversion of the following vector
\be\label{LR15aa} %
{\bf \tM}_n(s)=\frac{1}{s-A_n(s)}{\bf M}_n(0) \, ,  %
\ee %
where $(2n+1)\times(2n+1)$ matrix $A_n(s)$ consists of
coefficients from the matrix equation (\ref{LR9}). In the limit
$s\rightarrow 0$ the disorder term of order of
$s^{\alpha}\rightarrow\infty$ is dominant, and the maximal
eigenvalues of $A_n$ can be evaluated at the energy $\ep\approx
0$.  Following Ref. \cite{fik}, it can be proven  that for $\ep=0$
the maximal eigenvalues of $A_n$ behaves for large $n$ as
$\Omega(s)\approx s^{\alpha/3}(2n)^{4/3}$. Expanding the initial
condition $M_n(0)$ over the eigenfunctions of $A_n$, we obtain
that the maximal growth of the $n$th moment is \be\label{LR15ab}
M_n(t)=
\LT^{-1}\Big[\frac{s^{-\alpha/3}}{s^{1-\alpha/3}-(2n)^{4/3}}\Big]M_{\Omega}(0)\,
. \ee The inverse Laplace transform is the definition of the
Mittag-Leffler function \cite{BE}:
$E_{1-\alpha/3}\Big(\frac{3}{4}(2n)^{4/3}t^{1-\alpha/3}\Big)$.
Asymptotic behavior of the Mittag-Leffler function for
$t\rightarrow\infty$ is determined by the exponential function
$\exp\Big[(2n)^{4/(3-\alpha)}t\Big]$. Therefore the exponential
growth of the $n$th moment is due to the Lyapunov exponent
\be\label{LR18_a} %
\gamma(0)\sim (2n)^{4/(3-\alpha)} %
\ee %
for $-1<\alpha<0$.

For $\alpha >0$ the fractional integral diverges. To overcome this
obstacle, one considers the causal function as a generalized
function, and a suitable regularization procedure can be carried
out  see e.g., \cite{klafter,zaslavsky}. Let $N-1<\alpha<N$, where
$N\geq 1$ is an integer. Again, using the composition rule, one
obtains the Riemann-Liouville fractional integral (\ref{LR10}) in
the regularized form
\bea\label{LR11} %
D_t^{\alpha} f(t)&=&D_t^ND_t^{\alpha-N}f(t) \equiv
D_{RL}^{\alpha}f(t)  \nonumber \\
&=&
\frac{1}{\Gamma(N-\alpha)}\frac{d^N}{dt^N}\int_0^t
\frac{f(\tau)d\tau}{(t-\tau)^{1+\alpha-N}} \, .
\eea %
Thus Eq. (\ref{LR9}) reads
\be\label{LR13} %
\dot{M}_{k,l}=kM_{k+1,l-1}-l\ep M_{k+1,l-1}+
l(l-1)D_{RL}^{\alpha}M_{k+2,l-2} \, . %
\ee  %
This fractional equation of the order of $\alpha$ must be equipped
with $N-1$ quasi initial conditions: in addition to the initial
conditions $M_{k,l}(0)$, one has to know $N-1$ fractional
derivative of $M_{k,l}(\tau)$ at $\tau=0$. Application of the
Laplace transform to the fractional derivative yields \cite{klafter}
\be\label{LR14} %
\LT
[D_{RL}^{\alpha}M_{k,l}(t)]=s^{\alpha}\tM_{k,l}(s)-\sum_{p=0}^{N-1}
s^pD_{RL}^{\alpha-1-p}M_{k,l}(t)\Big|_{t=0}\, . %
\ee %
In the asymptotic limit $s\rightarrow 0$ we obtain that the
solution of Eq. (\ref{LR13}) is approximated by the inverse
Laplace transform of the vector
\be\label{LR15} %
{\bf \tM}_n(s)\approx\frac{1}{s-A_n(s)}[I_n-B_nD_{RL}^{\alpha-1}]{\bf M}_n(0) \, ,  %
\ee %
where $I_n$ is an unit matrix and matrix $B_n$ consists of the off diagonal elements which produce
$M_{k+2,l-2}$ terms in Eq. (\ref{LR13}).

Since we are seeking the maximal growth rate of the solution of
Eq. (\ref{LR13}) and the initial condition are not important for
this growth, we choose the initial condition as the eigenvector of
the maximal eigenvalue of the matrix $A_n$. In what follows we
consider a temporal behavior of the second moments, described by
$3\times 3$ matrix $A_1(s)$. The eigenvalues of the matrix are
roots of a cubic equation \cite{if07}. The growth rate is
determined by the eigenvalue with the largest real part that will be
denoted by $\Omega$. Taking the initial condition in Eq.
(\ref{LR15}
 as the eigenfunction of $\Omega$, namely $-{\bf M}_{\Omega}(0)$, we obtain that the dynamics of
the second moments is due to the Laplace inversion
\be\label{LR16} %
{\bf M}_1(t)\propto\LT^{-1}\Big[(\Omega(s)-s)^{-1}\Big]\, . %
\ee  %
For small $s$ the eigenvalues $\Omega(s)$ correspond to a ``weak''
disorder in the Laplace space. Therefore the high energy limit is
valid $\Omega\approx s^{\alpha}/\ep$, where $\ep\gg s^{\alpha}$
\cite{LGP,if07}. Substituting this eigenvalue in Eq. (\ref{LR16})
and expanding  the denominator we have for the integrand
$\sum_{n=0}^{\infty}\ep^{n-1}(1/s)^{(\alpha-1)n+\alpha}$. Carrying
out the Laplace inversion, we obtain the solution in the form of
another definition of the Mittag-Leffler function (see e.g.,
\cite{klafter,podlubny}) $ E_{\alpha-1,\alpha}(\ep t^{\alpha-1})=
\sum_{n=0}^{\infty} \frac{(\ep
t^{\alpha-1})^n}{\Gamma(n\alpha-n+\alpha)}$. Therefore
\be\label{LR17} %
{\bf M}_1(t)\propto \ep t^{\alpha-1} E_{\alpha-1,\alpha}(\ep t^{\alpha-1})\, .
\ee  %
Since the argument of the Mittag-Leffler function is positive $\ep
t^{\alpha-1}>0$, then the asymptotic behavior is approximately
$E_{\beta,\delta}(z)\approx z^{(1-\delta)/\beta}\exp(z^{1/\beta})$
for $z\rightarrow\infty$ for all values $\delta$ \cite{BE,podlubny}.
Therefore, when $\ep t^{\alpha-1}\rightarrow\infty$ the
exponential growth of the second moment
\be\label{LR18_aabb}
{\bf M}_1(t)\propto \exp[\gamma(\ep) t]  %
\ee %
is approximated by the Lyapunov exponent
\be\label{LR18_b} %
\gamma(\ep)\sim \ep^{1/(\alpha-1)}\, . %
\ee %

Another way to obtain the Lyapunov exponents avoiding the
difficulties related to the $N-1$ quasi initial conditions in Eqs.
(\ref{LR13}) and (\ref{LR14}) is to discard the causality
principle and extend the consideration of the random process on
the entire $x$ axis $x\in(-\infty,+\infty)$. For this formal
consideration, the Furutsu-Novikov formula in Eq. (\ref{LR8})
reads \bea\label{LR19}
-(l-1)C_{\alpha}\int_{-\infty}^x\frac{M_{k+2,l-2}(y)}{(x-y)^{1+\alpha}}dy
~~~\mbox{for $x>0$}\, ,\nonumber \\
-(l-1)C_{\alpha}\int_{x}^{\infty}\frac{M_{k+2,l-2}(y)}{(y-x)^{1+\alpha}}dy
~~~\mbox{for $x<0$}\, . \eea Setting again
$C_{\alpha}=\Gamma(-\alpha)$, we obtain that Eq. (\ref{LR19}) is
the definition of the Riesz/Weyl fractional derivative
$\clW_x^{\alpha}$ see e.g., \cite{klafter,zaslavsky,podlubny}.
Therefore, Eq. (\ref{LR9}) now reads
\be\label{LR20}
\frac{d}{dx}M_{k,l}=kM_{k-1,l+1}+l\ep
M_{k+1,l-1}+l(l-1)\clW_x^{\alpha}M_{k+2,l-2}\, .
\ee
A specific
property that we use is the fractional differentiation of an
exponential $\clW_x^{\alpha}\exp(\gamma x)=
\gamma^{\alpha}\exp(\gamma x)$. Substituting this in Eq.
(\ref{LR7}), one seeks the solution for the maximal moment growth
$M_{k,l}(x)=\exp(\pm\gamma x)M_{k,l}(x=0)$, where plus stays for
$x>0$ and minus for $x<0$, respectively. One readily checks that
the both cases yield the same algebraic equation
\be\label{LR21} %
\gamma{\bf M}_n=A_n(\gamma){\bf M}_n\, , %
\ee  
where the moment vector ${\bf M}_n$ is defined above and the
matrix $A_n(\gamma)$ is defined from Eq. (\ref{LR20}). Therefore,
$\Omega(\gamma)=\gamma^{\alpha}/\ep$, where conditions $\gamma\ll
\ep$ and $\gamma^{\alpha}\ll \ep$ are used. Solutions of Eq.
(\ref{LR20})  for $\gamma(\ep)$ coincide exactly with the ones
obtained in Eqs. (\ref{LR18_a}) and (\ref{LR18_b}) for all values
of $\alpha$.

This solution for $\gamma$ also yields conditions of validity of
the solution (\ref{LR18_b}) for different values of energy $\ep$.
Indeed, for $0<\alpha<1$ Eqs. (\ref{LR18_aabb}), (\ref{LR18_b})
and (\ref{LR21}) describe an exponential growth for asymptotically
large energies $\ep\gg 1$, since, in this case,
$\gamma(\ep)\ll\ep$ when $\ep\gg 1$.
 On the contrary, when $\alpha>1$  the solution of Eq.
(\ref{LR18_b}) is valid for $\ep\ll 1$. This follows from the
condition $\gamma^{\alpha}\ll\ep$. Note that for large negative
values of the energy $\Omega\sim 2\sqrt{|\ep|}$, what corresponds
to a simple pole in Eq. (\ref{LR16}), and this is just the Lyapunov
exponent $\gamma(\ep)\sim 2\sqrt{|\ep|}$ .

In conclusion, we studied the Lyapunov exponents for Anderson
localization in a one-dimensional disordered system with a
long-range memory. The averaged behavior of the second moment of
the eigenfunction is calculated, and its asymptotic exponential
growth for $|x|\rightarrow\infty$ is determined by the Lyapunov
exponents for different values of the energy $\ep$. The main
result of the study is the existence of the positive Lyapunov
exponents $\gamma(\ep)>0$ for the rate $q=1+\alpha>0$ of the power
law decay of the correlation function. It is relevant to the
exponential localization of the eigenfunctions of the random
Schr\"odinger operator of Eq. (\ref{LR1}).

This work was supported by the Israel Science Foundation. I thank
S. Fishman for very informative and instructive discussions.


\begin{thebibliography}{11}

\bibitem {Anderson}P.W. Anderson, Phys. Rev. {\bf 109}, 1492
(1958).

\bibitem{Lee} P. A. Lee and T. V. Ramakrishnan, Rev. Mod.
Phys. {\bf 57}, 287 (1985).

\bibitem{G13} L. Sanchez-Palencia et al., Phys. Rev. Let. {\bf
98}, 210401 (2007); J. Billy et al., Nature {\bf 453}, 891 (2008).

\bibitem{G14} G. Roati  et al., Nature {\bf 453}, 895 (2008).

\bibitem{israilev} U. Kuhl et al., Appl. Phys. Let. {\bf 77}, 633
(2000); U. Kuhl and H.-J. St\"ockmann, Phys. Rev. lett. {\bf
80},3232 (1998).

\bibitem{garcia} A.M. Garcia-Garcia and E. Cuevas,
Absence of localization in one-dimensional disordered systems,
cond-mat.0808.3757.

\bibitem{moura} F.A.B.F. de Moura and M.L. Lyra, Phys. Rev. Let.
{\bf 81}, 3735 (1998).

\bibitem{LyraHavlin} F.A.B.F. de Moura and M.L. Lyra, Physica
A {\bf 266}, 465 (1999); S. Russ, et al., Physica A {\bf 266}, 492
(1999).

\bibitem{G7} S. Kotani, Proc.Kyoto Stoch. Com. (1982); B. Simon,
Comm. Math. Phys. {\bf 89}, 227 (1983).

\bibitem{G8} S. Kotani and B. Simon, Comm. Math. Phys. {\bf 112}, 103
(1987).

\bibitem{PF} L.A. Pastur and A.L. Figotin, \textit{Spectra of Random
and Almost-Periodic Operators} (Springer, Berlin, 1992).

\bibitem{LGP} I.M. Lifshits, S.A. Gredeskul, and L.A.
Pastur, \textit{Introduction to the theory of disordered systems}
(Wiley-Interscience, New York, 1988).

\bibitem{kirone} K. Mallick and P. Marcq, Phys. Rev. E {\bf 66},
041113 (2002).

\bibitem{schomerus} H.~Schomerus and M.~Titov,
  Phys. Rev. E {\bf 66}, 066207 (2002).

\bibitem{pikovsky} R. Zilmer and A. Pikovsky, Phys. Rev. E {\bf
67} 061117 (2003).

\bibitem{fik} S. Fishman, A. Iomin, and K. Mallick, Phys. Rev. E
{\bf 78}, 066605 (2008).

\bibitem{halperin} B.I. Halperin, Phys. Rev. {\bf 139}A, 104
(1965).

\bibitem{klyatskin} V.I. Kliatskin, \textit{Stochastic equations and
waves in randomly inhomogeneous media} (Nauka, Moskva, 1980) (in
Russian).

\bibitem{bouchaud} J.-P. Bouchaud and A. Georges, Phys. Rep. {\bf
195}, 127 (1990).

\bibitem{klafter} R. Metzler and J. Klafter, Phys. Rep. {\bf 339}, 1
(2000).

\bibitem{zaslavsky} G.M. Zaslavsky, Phys . Rep. {\bf 371}, 461 (2002).

\bibitem{BE} H. Bateman and A. Erd\'elyi \textit{Higher Transcendental functions}
(Mc Graw-Hill, New York, 1955), V. 3.

\bibitem{if07} A. Iomin and S. Fishman Phys. Rev. E {\bf 76}, 056607 (2007).

\bibitem{podlubny} I. Podlubny, \textit{Fractional Differential
Equations} (Academic Press, San Diego, 1999).

\end{thebibliography}
\end{document}